# A simple model for vacancy order and disorder in defective half-Heusler systems

Nikolaj Roth[a], Tiejun Zhu[b] and Bo Brummerstedt Iversen*[a]

Defective half-Heusler systems $X_{1-x}YZ$ with large amounts of intrinsic vacancies, such as $Nb_{1-x}CoSb$, $Ti_{1-x}NiSb$ and $V_{1-x}CoSb$, are a group of promising thermoelectric materials. Even with high vacancy concentrations they maintain the average half-Heusler crystal structure. These systems show high electrical conductivity but low thermal conductivity arising from an ordered YZ lattice, which conducts electrons, while the large amounts of vacancies on the X sublattice effectively scatters phonons. Using electron scattering it was recently observed that in addition to Bragg diffraction from the average cubic half-Heusler structure, some of these samples show broad diffuse scattering indicating short-range vacancy order while other samples show sharp additional peaks, indicating long-range vacancy ordering. Here we show that both the short and long-range ordering can be explained using the same simple model, which assumes that vacancies on the X-sublattice avoid each other. The samples showing long-range vacancy order are in agreement with the predicted ground-state of the model, while short-range order samples are quenched high-temperature states of the system. A previous study showed that changes in sample stoichiometry affect whether the short or long-range vacancy structure is obtained, but the present model suggests that thermal treatment of samples should allow controlling the degree of vacancy order, and thereby the thermal conductivity, without changes in composition. This is important as the composition also dictates the amount of electrical carriers. Independent control of electrical carrier concentration and degree of vacancy order should allow further improvements in the thermoelectric properties of these systems.

## Broader Context

Thermoelectric materials, which can convert waste heat into useful electrical energy, are promising for use in sustainable energy technology. To optimize thermoelectric efficiency, semiconductors with high electrical conductivity but low thermal conductivity are needed. High electrical conductivity is usually found in ordered crystalline materials while low thermal conductivities are associated with disordered and amorphous materials. Materials consisting of a conducting lattice with a disordered sublattice are therefore good candidate thermoelectric materials. Recently defective half-Heusler systems with large amounts of intrinsic atomic vacancies have been studied. From electron scattering measurements it has been shown that some of these systems have short-range vacancy order while others show signs of long-range vacancy ordering. Here we develop a simple model which can explain the observed electron scattering measurements. The model assumes that vacancies tend to avoid each other. This leads to a ground-state which is in agreement with the observed scattering data for the long-range vacancy order samples. Furthermore, the samples which only show signs of short-range vacancy order can be explained as quenched non-ground-state structures. This finding suggests that the degree of short-range and long-range order of the system can be controlled through thermal treatment of samples, allowing independent tuning of electrical and thermal conduction and therefore delicate fine-tuning of thermoelectric properties.

## Introduction

To combat climate change while supplying a growing world population with affordable and clean energy, improvements in sustainable energy production are needed[1]. One of several technologies useful to obtain such a goal is thermoelectric energy conversion, which allows direct conversion of waste heat into electrical energy[2]. To do so efficiently thermoelectric materials must have a high electrical conductivity together with a low thermal conductivity[3]. High electrical conductivities are usually found in highly ordered crystalline materials, while low thermal conductivities usually are associated with disordered or amorphous materials. One way to overcome this apparent paradox is though materials that have a long-range crystalline ordered lattice with a disordered sublattice. The ordered lattice should be responsible for the conduction of electrons while the disordered sublattice will efficiently scatter phonons. Such materials are typically understood using the phonon-glass electron crystal concept[3,4]. Examples include clathrates and skutterudites where conducting crystalline network-structures have additional guest atoms in loosely bonded positions, typically large voids. This allows the guest atoms to "rattle" independently of the network and thereby scatter phonons

[a.] Center for Materials Crystallography (CMC), Department of Chemistry and Interdisciplinary Nanoscience Center (iNANO), Aarhus University, Langelandsgade 140, Aarhus 8000, Denmark
[b.] State Key Laboratory of Silicon Materials, School of Materials Science and Engineering, Zhejiang University, Hangzhou 310027, China

efficiently. Another example is $Cu_{2-x}Se$, which consists of a long-range ordered Se lattice allowing electronic transport, while the Cu-sublattice is disordered. At room-temperature the Cu atoms are ordered in two-dimensions but disordered along the third dimension, and above about 400K the Cu sublattice becomes almost liquid-like[5,6]. To further understand such phonon-glass electron crystal materials accurate knowledge of their atomic structure is required.

Ordered crystalline materials give rise to sharp Bragg peaks in a scattering experiment. Conventionally, the structures of crystalline materials are solved by analyzing the Bragg peak intensities from either x-ray or neutron scattering experiments. The sharp Bragg peaks contain information about the unit-cell average structure. For perfectly ordered crystals this is the same as the complete structure. However, for crystals containing disorder this is no longer the case. For crystals with large movements of atoms, such as the clathrates, the average structure only shows the smeared-out average electron density of the moving atoms. For crystals with substitutional disorder only the average electron-density of a site can be known from analysis of Bragg intensities, and all information about local correlations is lost in the average structure.

To obtain information about the local correlations it is necessary to analyze the weak diffuse scattering, which contains information about disorder and short-range order[7-10]. In crystals with correlated movements of atoms, the diffuse scattering can show the actual instantaneous configuration of atoms. A recent example relevant to thermoelectric research is the existence of local dynamic dipole-formation in PbTe[11]. In crystals with substitutional or vacancy disorder the diffuse scattering contains information about the short-range order structure. As an example, this was recently used to show that thermoelectric $Cu_{2-x}Se$ at room-temperature contains ordered two-dimensional layers of Cu, whereas the structure is disordered along the third dimension[6].

Half-Heusler compounds with general composition XYZ, such as ZrNiSn, ZrCoSb and NbFeSb, among others, have been intensively studied for use as thermoelectric materials due to their favorable properties[12-20]. These compounds all have a valence electron count of 18, which for the half-Heusler structure results in filling of all bonding electronic states, giving a highly stable structure. As these compounds have all states filled up to the band-gap, they behave as semiconductors. The main drawback of these compounds has been their relatively high lattice thermal conductivities, which result from their simple crystal structure, although nanostructuring and alloying have been applied to reduce this issue[21,22]. Recently a new group of nominal 19-electron half-Heusler compounds have been studied, e.g. VCoSb, NbCoSb and TiNiSb[23-28]. With the extra electron compared to 18-electron half-Heusler compounds, these new compounds should be expected to be metallic and less stable. However, instead they have been found to behave as strongly doped semiconductors with much lower thermal conductivities than the 18-electron systems[23-27], making them very suitable for thermoelectric applications. The explanation for this is that the systems in fact have large deficiencies of the X element, getting them closer to an 18-electron composition. As an example the nominal 19-electron NbCoSb was shown to consist of a phase with a composition close to the 18-electron $Nb_{0.8}CoSb$, and additional Nb-rich impurity phases. The $Nb_{1-x}CoSb$ phase has been reported with stoichiometry $Nb_{0.84}CoSb$ by Zeier et al. and $Nb_{0.79}CoSb$-$Nb_{0.83}CoSb$ by Xia et al.[25,26]

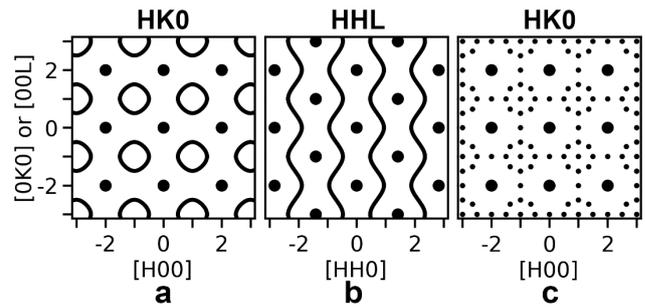

**Fig. 1** Sketch showing the scattering measured for a number of defective half-Heusler systems by Xia et al. (ref 27) (a) and (b) show the HK0 and HHL plane for short range order systems $Nb_{0.8}CoSb$, $Ti_{0.9}NiSb$, $V_{0.9}CoSb$ and $Nb_{0.8}Co_{0.92}Ni_{0.08}Sb$, (c) shows the HK0 plane for ordered $Nb_{0.82}CoSb$ and $Nb_{0.84}CoSb$. Large dots indicate the reflections from the average structure, lines indicate diffuse scattering and small dots indicate weak additional peaks.

These new compounds are therefore better described as defective half-Heusler $X_{1-x}YZ$ systems. Interestingly, these compounds still maintain the average cubic half-Heusler crystal-structure, even though they often have around a 20% deficiency of the X element[25,26]. Their electronic structure can be explained using the 18-electron rule, where the additional/fewer Nb atoms compared with $Nb_{0.8}CoSb$ will act as heavy dopants[25].

Because of impurities formed during synthesis additional characterization of these compounds is needed to know the actual stoichiometry of the half-Heusler phases. The sample with nominal composition NbCoSb produced by Zeier et al. was found to have a composition of about $Nb_{0.84}CoSb$ using Rietveld-refinement of powder X-ray diffraction data[25]. Using an Electron probe micro-analyzer wavelength dispersive spectroscope Xia et al. measured the stoichiometries of the half-Heusler phases in their samples [26]. For all phases with nominal stoichiometry of $Nb_{1-x}CoSb$ in the range x=0-0.25 they found that the stoichiometry of the half-Heusler phase only vary in the interval x=0.17-0.21. For nominal compositions outside this region, additional impurity phases were observed.

With this large number of missing Nb atoms, the question arises of how these vacancies are distributed. Zeier et al. suggested an ordered tetragonal vacancy structure for $Nb_{0.8}CoSb$ based on DFT calculations[25]. Recently Xia et al. investigated the electron scattering from several of these defective half-Heusler systems[27]. For $Nb_{0.8}CoSb$, $Ti_{0.9}NiSb$, $V_{0.9}CoSb$ and $Nb_{0.8}Co_{0.92}Ni_{0.08}Sb$ similar scattering patterns were observed showing both strong Bragg peaks from the average cubic half-Heusler structure and weak diffuse scattering coming from short-range order of vacancies. In the HK0 plane of the scattering, diffuse rings were found around the systematically extinct reflections with both odd h and k. This is sketched in

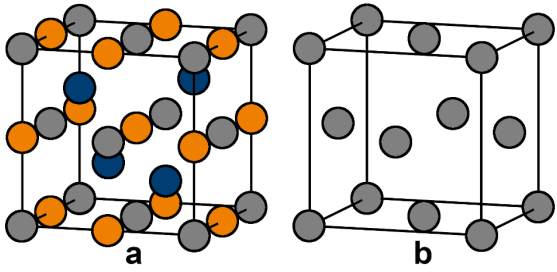

**Fig. 2** Crystal structure of the defective $X_{1-x}YZ$ half-Heusler systems. (a) Shows the full structure with X (grey), Y (blue) and Z (orange). (b) The X/vacancy fcc sublattice.

figure 1a. In the HHL plane (spanned by HH0 and 00L) continuous waves of diffuse scattering were observed between rows of Bragg peaks along the 00L direction, as sketched in figure 1b. In the $Nb_{1-x}CoSb$ system Xia et al. also observed several ordered structures depending on the nominal stoichiometry of the samples[27]. For the $Nb_{0.81}CoSb$ sample the scattering in the HK0 plane showed two additional peaks on the diffuse scattering ring on opposite sides of the ring along one of the [H00] directions, and further peaks were found half-way between the main reflections of the structure in the same direction. For the samples with nominal stoichiometry $Nb_{0.82}CoSb$ and $Nb_{0.84}CoSb$ (measured stoichiometry $Nb_{0.81}CoSb$ and $Nb_{0.82}CoSb$[26]) the circles of diffuse scattering were mainly gone but replaced by an 8-point ring, with additional very weak peaks in the center of the ring and between main reflections along all [100] directions, sketched in figure 1c.

As very similar types of short range order diffuse scattering have been observed in several different half-Heusler systems it seems to be a common feature of these systems. Here we give a simple model for both the diffuse scattering as well as the ordered structures observed by electron diffraction by Xia et al. The model only considers the positions of vacancies on the X-sublattice of the structure.

## Theory

The average structure of the defective $X_{1-x}YZ$ half-Heusler systems is shown in figure 2a. The half-Heusler structure has X on a face-centered cubic lattice with Z in the octahedral positions and Y in half of the tetrahedral positons. X and Y together form the sphalerite structure while X and Z together form the rock salt structure. As these systems have a deficiency of X, there will be vacancies in the structure. It was found that there are no apparent anti-site defects in these systems[25] as there are for other half-Heusler systems[29,30]. The vacancies are therefore limited to the face-centered cubic sublattice of the half-Heusler structure, as shown in figure 2b.

The question of the atomic short and long range order can therefore be simplified to how the vacancies are distributed on the fcc lattice. To illustrate different vacancy distributions we will use a different representation of this structure, where the fcc sublattice is viewed from above as shown in figure 3a. Here the grey circles show the sites with whole integer z-coordinates, and the grey triangles show sites with half-integer z-coordinates. In this view, the unit cell axis are along the diagonal, as marked by the red square. This representation is convenient for representing the ordered vacancy structures as shown below.

We make the simple assumption that vacancies avoid each other, and only take into account nearest and next-nearest neighbor interactions. When there are more than 1/4 vacancies on the X sublattice it is not possible to arrange the vacancies such that there are no nearest-neighbors. When x=1/4 it is possible to make an ordered arrangement with no nearest-neighbors, but with each vacancy having 4 out of 6 next-nearest neighbors also vacant. This arrangement is illustrated in figure 3b. In order to both completely avoid nearest and next-nearest neighbor vacancies, 1/6 sites need to be vacant. Several possible arrangements will satisfy these conditions, and one such arrangement is shown in figure 3d. More are shown in the supporting information. This means that for x<1/4 nearest neighbors can be eliminated and for x<1/6 also all next-nearest neighbors can be eliminated. In cases where 1/4>x>1/6 there has to be some next-nearest neighbor vacancies if nearest-neighbor vacancies are still to be avoided.

The ordered vacancy structure for x=1/5 suggested by Zeier et al.[25] is illustrated in figure 3c. In this structure, each vacancy has no nearest-neighbors but two next-nearest neighbors.

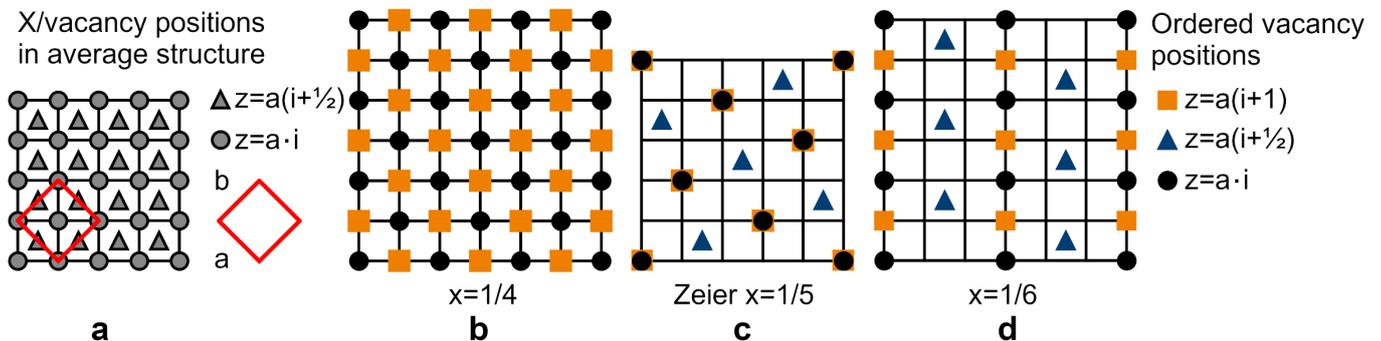

**Fig. 3** Vacancy positions on the fcc lattice as seen from above. (a) The possible vacancy positions. Grey circles mark sites with integer z-coordinates and grey triangles mark sites with half-integer z-coordinates. The a and b axis of the unit cell are along the diagonal as shown by the red square. (b) One possible ordering of vacancies for x=1/4. In this ordering there are no vacancies with half-integer z-coordinates. This ordering has no nearest neighbor vacancies, and each vacancy has 4 next-nearest neighbor vacancies. (c) The vacancy ordering with x=1/5 suggested by Zeier et al. giving a tetragonal structure. Note that every second layer is identical. This ordering has no nearest-neighbor vacancies, but every vacancy has two next-nearest neighbor vacancies. (d) One of several possible vacancy orderings for x=1/6 where there are no nearest and no next-nearest neighbor vacancies. Other possible orderings are shown in the supporting information.

However, if instead of adapting the ordered x=1/5 structure the vacancy sublattice creates domains of x=1/6 and x=1/4 type structures there will be a lower average number of next-nearest neighbors. This will allow 3/5 of the vacancies to go into the x=1/6 structure, leaving 2/5 of vacancies in the x=1/4 structure. This will give an average number of next-nearest neighbors per vacancy of 1.6, somewhat lower than in the structure suggested in reference 25. This would suggest that domain-phase separation of the sublattice is more favorable for 1/4>x>1/6.

For x<1/6 there are many states all obeying no nearest and next-nearest neighbor vacancies and the model system will be disordered. To further investigate the vacancy distribution we simulate the system using Monte-Carlo simulations for different compositions, x, and temperatures. As we are only considering nearest and next-nearest neighbor interactions of the vacancies, the energy of the model system can be written as a sum over all $N_{vac}$ vacancies:

$$E = \frac{1}{2} \sum_{i=1}^{N_{vac}} \left( \sum_{j}^{12NN} S_{ij} + J' \sum_{j'}^{6NNN} S_{ij'} \right) \quad \text{(equation 1)}$$

Where the sum over j goes over the 12 nearest-neighbor (NN) sites of vacancy i. $S_{ij}$ is 1 if site j has a vacancy and 0 if it is occupied. Similarly the sum over j' is over the 6 next-nearest neighbors (NNN) of vacancy i where $S_{ij'}$ is 1 if site j' is vacant and 0 if occupied. In this formalism an energy of 1 is assigned to a nearest-neighbor vacancy pair and J' for a next-nearest neighbor vacancy pair.

## Methods

Monte-Carlo (MC) simulations were carried out using a custom code written in python. A box of 24*24*48 sites with the appropriate number of vacancies were started in either a random configuration or from one of the ordered states. In cases where the ordered starting state was used for a different x than that of the ordered state, either additional vacancies were added randomly or random vacancies were removed in order to have the correct number of total vacancies. Vacancies were allowed to move to nearest-neighbor positions following the Metropolis algorithm[31] with the energy expression given in equation 1. In cases where convergence was hard to reach, a simulated annealing approach was used, where the temperature of the simulation was continuously lowered. Each simulation was repeated 20 times to increase the statistics of the following scattering calculation. The simulations shown in this paper used J'=0.48 as the relative value for the next-nearest-neighbor energy. This value was chosen as a J' lower than 0.5 reproduces the observed ground-state scattering well while 0.45 or higher reproces the observed short-range order scattering at higher simulated temperatures well.

The scattering patterns of the configurations were calculated using the software scatty[32] with Nb on the positions of the occupied X sites. X-ray scattering factors were used.

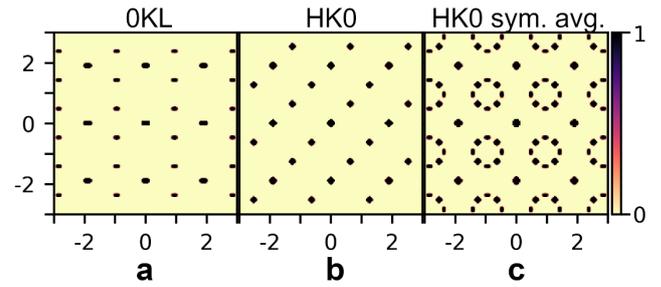

**Fig. 4** Calculated scattering from the ordered x=1/6 structure. (a) and (b) show the 0KL and HK0 plane which are equivalent in the average cubic structure. (c) The HK0 plane of the scattering symmetry averaged to the average structure, which would be expected for micro-domains with different but equivalent orientations.

As the average structure is cubic, while the proposed ordered vacancy structures are not, different scattering signal should be observed in directions, which are equivalent in the average cubic structure. As an example, the calculated scattering from the ordered x=1/6 model is shown in figure 4 a and b for the 0KL and HK0 planes, which are symmetry equivalent in the average cubic structure. The 0KL plane shows additional peaks corresponding to the ones Xia et al observed for their $Nb_{0.81}CoSb$ sample[27], although without the additional diffuse scattering they observed. The HK0 plane is different with the additional peaks forming rows along one diagonal. The H0L plane, not shown here, is equivalent to the 0KL plane. It could be expected that this type of system will have many micro-domains, where different regions of the sample will orient along the different equivalent directions of the average cubic structure. If the scattering of the ordered x=1/6 structure is symmetrized to the Laue symmetry of the average structure, the scattering pattern shown in figure 4c is obtained. Now the 8-point rings have emerged, which are very similar to the ones observed in several samples by Xia et al[27], however without the peaks at the center of the ring. This however shows that most of the reported samples have micro-domain structure with different orientations, which are symmetry-equivalent in the average cubic structure. The 8 points in the observed rings do therefore not all come from one domain, which would require an incommensurate modulated structure, but come in pairs from different orientations of a commensurate superstructure. In the following, we will focus on the cubic symmetry averaged scattering of the different phases, as these are the patterns most likely to be observed in an experiment where the sample will contain many micro-domains.

Figure 5 shows the scattering patterns for the ground state structures for different compositions. The figure shows both the HK0 (upper row) and HHL planes (lower row) which have been symmetry averaged to the Laue symmetry of the average cubic structure. The scattering from the ordered x=1/4 structure presented in figure 3b is shown in first column of figure 5. Characteristic of the HK0 plane are the rows of additional peaks along odd-integer H and K. The scattering of the ordered x=1/6 structure presented in figure 3d is shown the third column of

## Results

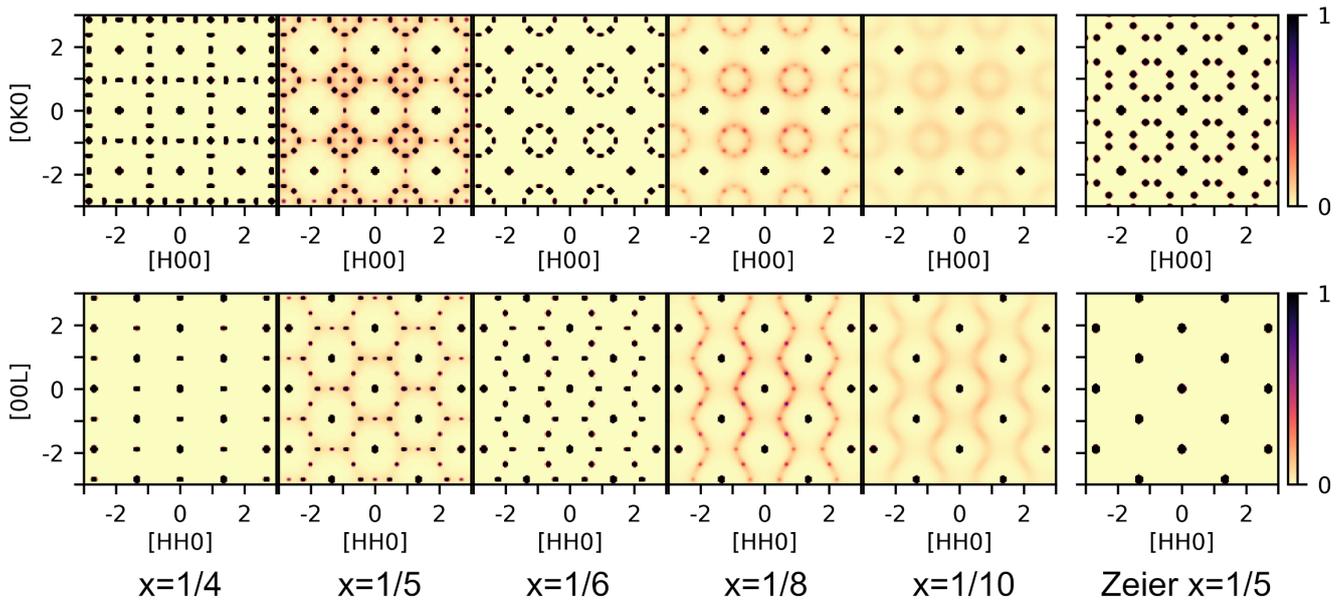

**Fig. 5** Calculated scattering for the ground states at different compositions in the HK0 (upper row) and HHL planes (lower row). For all these models the scattering has been averaged to the symmetry of the average structure to simulate the expected measurement for a sample with multiple domains. The scattering for x=1/4, x=1/6 as well as the x=1/5 structure suggested by Zeier et al. are calculated from the ordered vacancy structures shown in Fig. 3. For the x=1/5, x=1/8 and x=1/10 systems the scattering is calculated from the Monte-Carlo simulated structures. The calculated scattering for several other possible ground-state vacancy structures and the scattering in non-symmetry averaged planes are shown in the supporting information.

figure 5. As was shown before, the HK0 plane has the 8-point ring without a peak at the center. In the region 1/4>x>1/6 the structure obtained from Monte-Carlo simulation has scattering which has both features from the x=1/4 and x=1/6 structure with additional diffuse scattering. This is shown in the second column of figure 5 for a simulation with x=1/5. The 8-point rings here also have peaks at their center and there are additional peaks between the main reflections. In this model there are inclusions of the x=1/4 type regions in the x=1/6 type structure. The scattering of this model matches that reported by Xia et al. for samples with nominal stoichiometry $Nb_{0.82}CoSb$ and $Nb_{0.84}CoSb$ (measured stoichiometry $Nb_{0.81}CoSb$ and $Nb_{0.82}CoSb$ [26]), as sketched in figure 1c. The structure suggested by Zeier et al. for x=1/5 based on DFT calculations[25] has a different scattering pattern to what has been observed, as seen in the last column of figure 5. As suggested in the theoretical section of this paper, it might be more favorable to have separation into domains of the ordered x=1/4 and x=1/6 types than to adapt the structure suggested by Zeier et al., and indeed the scattering from the ordered structures measured by Xia et al corresponds to this model.

The scattering from a simulated structure with x=1/8 is shown in the fourth column of figure 5. Here weak peaks of the ordered x=1/6 structure remain on top of strong diffuse scattering. When x=1/10 there is only diffuse scattering left, as shown in the fifth column od figure 5. This diffuse scattering is in agreement with the scattering measured by Xia et al. on $Ti_{0.9}NiSb$ and $V_{0.9}CoSb$ samples[27]. The calculated scattering for several other possible ground-state vacancy structures, including non-symmetry averaged planes are shown in the supporting information.

So far only the ground states for different compositions have been discussed. However, during the simulations it was found that it was hard to reach the ground states when starting from a random vacancy distribution unless simulated annealing was employed, or if the simulation was started from one of the ordered states close to the disordered ground state. This could suggest that real samples might also have difficulties reaching equilibrium, and that they may be trapped in higher temperature configurations if quenched. Therefore, Monte-Carlo simulations were carried out to simulate the effect of temperature on the system.

To illustrate the effect of temperature on the system, the scattering calculated for different temperatures of a Monte-Carlo simulation for x=1/5 is shown in figure 6. The left column shows the 0KL plane without the symmetry averaging, while the middle and right rows show the HK0 and HHL plane, which have been symmetry-averaged to the Laue symmetry of the average cubic structure. The left column is therefore the expected signal for one of the HK0 planes for a single domain sample, while the middle column shows the expected measurement in the HK0 plane for a sample composed of several domains with different orientation. The bottom row is the Monte-Carlo ground state and the simulated temperature is higher for the middle and upper row. The single-domain HK0 plane scattering of the low-temperature simulation, shown in the lower left corner of figure 6, is very similar to the measured scattering for the $Nb_{0.81}CoSb$ sample reported by Xia et al.[27]

As the temperature of the simulation is increased, the long-range order disappears, leaving the short-range order. First the peaks corresponding to the inclusions of the ordered x=1/4 phase disappear, leaving peaks from the ordered x=1/6 phase together with an increased amount of diffuse scattering, as seen in the middle row of figure 6. Then at higher temperatures all long-range order peaks are lost, leaving the diffuse scattering of the short-range order state, shown in the upper row of figure 6. This high-temperature short-range order state has similar scattering to the short-range order $Nb_{0.8}CoSb$ sample measured by Xia et al[27], as sketched in figure 1 a and b. The same high-temperature short-range order scattering is also obtained when the stoichiometry is different, as shown in the supporting information.

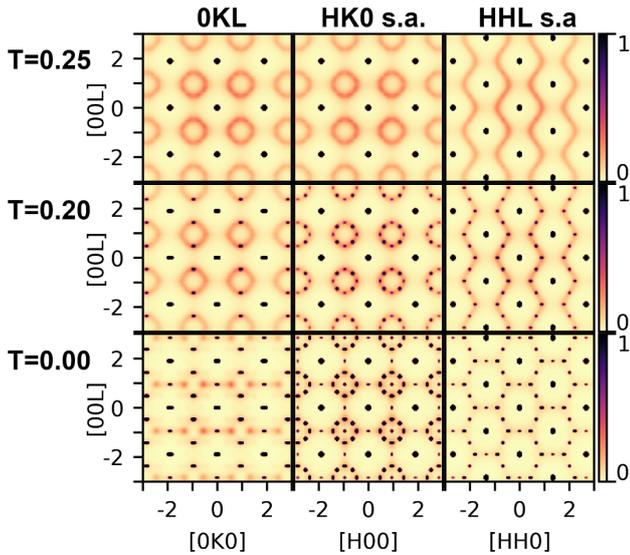

**Fig. 6** Effect of temperature on the scattering from the defective half-Heusler systems. Left column shown the 0KL plane, the middle column shows the cubic symmetry averaged HK0 plane and the right columns shows the symmetry averaged HHL plane at simulated temperatures T=0 (bottom row), T=0.2 (middle row) and T=0.25 (upper row). At increased temperatures the system loses its long-range order but retains short range order. The simulated temperature T is not the physical temperature but the parameter in the Monte-Carlo simulation which regulates how often positive-energy moves are accepted. The same high-temperature short-range order scattering is also obtained when the stoichiometry is different, as shown in the supporting information.

## Discussion

The scattering measured by Xia et al. on different samples can all be explained through the simple model presented above for vacancy distribution in the $X_{1-x}YZ$ systems. The short-range order scattering from $Ti_{0.9}NiSb$, $V_{0.9}CoSb$ matched the simulations with x=1/10, as shown in figure 5. The scattering from the sample with nominal stoichiometry $Nb_{0.81}CoSb$ (measured stoichiometry $Nb_{0.80}CoSb$[26]) matches the scattering from a single-orientation ground state with x=1/5, shown in the bottom left of figure 6. Scattering from samples with nominal stoichiometry $Nb_{0.83}CoSb$ and $Nb_{0.84}CoSb$ (measured to $Nb_{0.81}CoSb$ and $Nb_{0.82}CoSb$[26]) match the model scattering in the region 1/4>x>1/6 with multiple domains of the same type oriented along different directions, which are equivalent in the average cubic structure, shown in the second column of figure 5. The short-range order scattering from samples $Nb_{0.8}CoSb$ and $Nb_{0.8}Co_{0.92}Ni_{0.08}Sb$ match the high-temperature simulation for x=1/5, shown in the upper row of figure 6.

This shows that for $Nb_{1-x}CoSb$ samples in the range 1/4>x>1/6 some have the ground state structure while others have the higher-temperature short-range ordering structures. One possible explanation for this is that some of the samples might have been trapped in the high-temperature state during synthesis. Indeed the samples are quenched directly from a melt during the used synthesis route[33]. This makes it possible that some samples are cooled faster than others, or even that some regions of the same sample are cooled faster than other regions, which could explain why some samples show the high-temperature state while others the ground state. This would suggest that it could be possible to change between the short range order state and ground state by controlling the cooling rate of the synthesis, or through subsequent thermal treatment. Another explanation could be that when x becomes larger than 1/6, it becomes harder for the samples to reach the ground state as there are more vacancies in the structure interacting with each other. This could also explain why Xia et al. observed more diffuse scattering when going from x=1/6 towards x=1/5[27]. Possibly both are true, such that either having larger vacancy concentration or faster cooling rates would favor the short-range ordered state. This suggests that it could be possible to control the degree of vacancy order without changes in composition. This is important as the composition also dictates the amount of electrical carriers. Independent control of electrical carrier concentration and degree of vacancy order should allow further improvements in the thermoelectric properties of these systems.

It was also found that some of the samples show scattering from only one orientation of the ground state structure, while others showed scattering from multiple domains with different orientations, which are symmetry equivalent in the average half-Heusler structure. This suggests that it is possible to control the size of these micro-domains. Possibly this is also affected by the cooling rate.

The diffuse scattering patterns of the short range order states, as seen in figures 6 (top row) are similar to other previously studied systems such as the vacancy short range order in transition metal carbides with rock-salt structure, e.g. $MC_{1-x}$ and $MN_{1-x}$, with M=Ti,Nb,Ni,V[34,35], the distribution of Li and Fe in rock-salt structure α-LiFeO₂[36], oxygen and fluorine distribution in oxyflouride $K_3MoO_3F_3$ [37], short range order in $Mg_{1-x}Yb_{2x/3}\square_{x/3}S$ and related systems[38,39] among others [40]. What is common for these systems is that they have substitutional disorder, either of two types of elements/ions or of one element and vacancies on an fcc-based lattice, often a sublattice of the structure. Many of these systems have been explained in terms of a cluster model where the structure is built from octahedral clusters each with as close as possible stoichiometry to the sample. This local invariance of stoichiometry, often attributed to Linus Pauling, can in many of these systems be fulfilled without long range order. The result of the invariant octahedral clusters on an fcc lattice is that the diffuse scattering is limited to the curved surface satisfying[35,41,42]

$\cos(\pi h) + \cos(\pi k) + \cos(\pi l) = 0$     (equation 2)

The intersection of this surface with two planes are identical to the drawn lines in figure 1 a and b. If the octahedral cluster model is also to explain the scattering observed for the short range order in the half-Heusler systems, the octahedra in the fcc sublattice should be as close as possible to the average stoichiometry. An example of one such octahedra is the six side-centered atoms of the fcc unit cell. However, every site of the fcc lattice is contained in six octahedra. Those six octahedra are made from all nearest and next-nearest neighbors to the site. For $x \leq 1/6$ this model would imply that all octahedra have at most one vacancy. This is equivalent to all vacancies having no nearest or next-nearest neighbors as is the case for the ground states found in this study. For $1/4 > x > 1/6$ each octahedra would sometimes contain two vacancies in the cluster model. Although this is also true for the ground states found here the two are not equivalent, as the cluster model will allow some nearest-neighbors, while the found ground states do not have nearest neighbor vacancies, only some next-nearest neighbors in this range. However, the high-temperature states found here for $1/4 > x > 1/6$, for which the scattering is shown in figure 6 (top row), are equivalent to the octahedral cluster model.

Some previously reported systems also show similar types of ordering as the half-Heusler samples. De Ridder et al. observed that the alloy $Ni_4Mo$ changes structure when annealed[42]. One of the obtained phases shows an electron diffraction pattern which matches the one calculated here for the x=1/5 structure suggested by Zeier et al [25], shown in figure 5 (right column). They also found that after annealing a sample of $Ni_3Mo$ a diffraction pattern with the 8-point ring was obtained, similar to the ones shown here. These ordered systems can also be explained in terms of the octahedral cluster model[41,42] with the addition that certain configurations of the octahedra are more favorable that others, for example that vacancies on the same octahedra tend to be opposite. This is equivalent to vacancies not having nearest-neighbors, making the model equivalent to the ones found in this study.

The short range order observed in these half-Heusler systems therefore belongs to a more general type of short range order found in many systems with substitutional or vacancy disorder on fcc sublattices. Although these systems are chemically very different, the rule that local stoichiometry should be conserved lead to the same topology of the diffuse scattering. Indeed there are many such families of equivalent short range order found in chemically very diverse systems[8]. The link between local rules and the topology of diffuse scattering is very interesting. The approach of using the invariant cluster models was developing already in 1974[41]. A general approach to extinction rules in diffuse scattering based on symmetry was developed by Withers et al.[43]. The diffuse scattering for a large number of networks made from invariant building blocks obeying simple rules was shown by Overy et al.[44]. Recently a new way of treating diffuse scattering has emerged, describing the structures using a modulation wave approach[45]. This method lends itself very well to describing types of systems where diffuse scattering is limited to specific surfaces, such as the case found here. Another very promising approach is to describe the diffuse scattering using a disordered superspace model for the structure, which allows to describe seemingly very complex systems using only few parameters[46].

## Conclusions

Defective half-Heusler systems $X_{1-x}YZ$ with large amounts of vacancies have previously been found to show interesting scattering patterns, indicating short-range order in some samples and long-range order in others. Here, we have shown that all observed scattering patterns can be explained in terms of a simple model. The model assumes that vacancies tend to avoid being close to each other. The vacancies are distributed on a face-centered cubic sublattice of the half-Heusler structure. By avoiding nearest neighbor vacancy pairs and minimizing next-nearest neighbor vacancy pairs, the observed ordered phases can be explained. Furthermore, by simulating the effect of temperature on the system, the observed short-range order phases can be explained as quenched non-ground-state phases. It is further found that some of the reported measurements come from a single orientation of the ground-state phase, while other reported measurements originate from several ground-state domains with different orientations.

As the observed short-range order state can be explained through this model as a quenched high-temperature state, while the observed ordered phase is the ground-state of the same system, it suggests that the occurrence of the two phases can be controlled through thermal treatment of samples. Such control of the degree of disorder/order in the system should in turn allow tunable thermoelectric properties, as the thermal conductivity is expected to depend on the degree of long-range order. It has previously been observed that changes in sample composition can affect whether short or long range vacancy ordering is obtained. However, the composition also dictates the number of electrical carriers. This study therefore suggest that the degree of vacancy order, and thereby the thermal conductivity, can be controlled independently of carrier concentration using thermal treatment, allowing further optimization of thermoelectric properties.

## Conflicts of interest

There are no conflicts to declare.

## Acknowledgements

Kasper Tolborg and Kristoffer Holm are thanked for fruitful discussions. This work was supported by the Villum Foundation. T.J. Z wants to thank the financial support from the National Science Fund for Distinguished Young Scholars (No. 51725102) and the Natural Science Foundation of China (No. 51761135127).